\documentclass[a4paper,11pt]{article}

\usepackage{amssymb,amsmath}
\usepackage{mathrsfs}

\allowdisplaybreaks \numberwithin{equation}{section}
\newtheorem{thm}{Theorem}[section]

\newtheorem{lem}[thm]{Lemma}
\newtheorem{defn}[thm]{Definition}

\newtheorem{example}[thm]{Example}

\newtheorem{remark}[thm]{Remark}

\newenvironment{rmk}{\begin{remark} \rm }{\end{remark}}
\newenvironment{prf}{{\it Proof.}}{\hfill$\Box$}

\newcommand\od{\mathrm{d}}

\newcommand{\nn}{\nonumber}

\newcommand\pd{\partial}

\newcommand{\om}{\omega}

\newcommand{\dt}{\delta}

\newcommand{\res}{\mathrm{res}}

\newcommand\C{\mathbb{C}}

\newcommand\Z{\mathbb{Z}}
\newcommand\Zop{\mathbb{Z^{\mathrm{odd}}_+}}

\newcommand\bA{\mathbf{A}}

\newcommand\bX{\mathbf{X}}
\newcommand\cA{\mathcal{A}}
\newcommand\cD{\mathcal{D}}
\newcommand\cF{\mathcal{F}}

\newcommand\cM{\mathcal{M}}
\newcommand\cN{\mathcal{N}}
\newcommand\cP{\mathcal{P}}
\newcommand\cU{\mathcal{U}}
\newcommand\cV{\mathcal{V}}

\newcommand\fD{\mathfrak{D}}
\newcommand\fg{\mathfrak{g}}

\newcommand\ra{\rangle}
\newcommand\la{\langle}

\begin{document}
\title{Bihamiltonian Structure of the Two-component Kadomtsev-Petviashvili Hierarchy of type B}
\author{Chao-Zhong Wu\thanks{wucz05@mails.tsinghua.edu.cn},\quad
Dingdian Xu\thanks{xudd06@mails.tsinghua.edu.cn}
\\
{\small Department of Mathematical Sciences, Tsinghua University, }\\
{\small Beijing 100084, P. R. China}}
\date{}
\maketitle

\begin{abstract}
We employ a Lax pair representation of the two-component BKP
hierarchy and construct its bihamiltonian structure with $R$-matrix
techniques.

 \vskip 1ex \noindent{\bf Key words}: BKP hierarchy,
Hamiltonian structure, $R$-matrix
\end{abstract}


\section{Introduction}

The Kadomtsev-Petviashvili (KP) hierarchy of type B (BKP for short)
was introduced in~\cite{DKJM-KPBKP,DKM-BKP}, and generalized to
multi-component cases by Date, Jimbo, Kashiwara,
Miwa~\cite{DJKM-KPtype} in the form of bilinear equations. Among
these multi-component integrable systems, the two-component BKP
hierarchy is of special interest.

In the original definition of the two-component BKP hierarchy, the
solution space of tau functions can be regarded as the vacuum orbit
in the two-component neutral free fermionic Fock representation of
the infinite dimensional Lie algebra $D_\infty$~\cite{DJKM-reduce,
JM}, which corresponds to the infinite Dynkin diagram of type
D~\cite{Kac}. The Lie algebra $D_\infty$ can be reduced to the
affine Lie algebra $D_n^{(1)}$ under the so-called
$(2n-2,2)$-reduction in \cite{DJKM-reduce}, see also \cite{JM, tv}.
This reduction reduces the two-component BKP hierarchy to a
hierarchy that is equivalent with the Kac-Wakimoto hierarchy
corresponding to the principal vertex operator realization of the
basic representation of $D^{(1)}_n$, the Drinfeld-Sokolov hierarchy
associated to the Lie algebra $D^{(1)}_n$ and the zeroth vertex
$c_0$ of its Dynkin diagram, as well as the Givental-Milanov
hierarchy satisfied by the total descendant for the $D_n$
singularity, see \cite{DS, FGM, GM, KW, LWZ, Wu} and references
therein. Such a reduction is analogous to the one that reduces the
KP hierarchy to the $n$th Gelfand-Dickey hierarchy (see e.g.
\cite{Dickey}) that corresponds to the reduction of Lie algebras:
$A_\infty\mapsto A^{(1)}_n$. So in this sense to compare the
two-component BKP hierarchy with the KP hierarchy would deepen our
understanding of integrable hierarchies and relevant theories, such
as Jacobi/Prym varieties in algebraic geometry and Landau-Ginzburg
Models of topological strings, see e.g. \cite{Sh, Ta, Ta-2BKP}.

In this article our aim is to study the two-component BKP hierarchy
from the view point of Hamiltonian structures. To our best
knowledge, this topic has not been considered in the literature,
possibly for the reason that the KP-analogue Lax pair representation
of the two-component BKP hierarchy was unknown. Recall that the
two-component BKP hierarchy was defined to be the bilinear equation
of a single tau function:
  \begin{align}\label{2BKP-bltau}
&\res_z z^{-1} X(\mathbf{t};z)\tau(\mathbf{t},\hat{\mathbf{t}})
X(\mathbf{t}';-z)\tau(\mathbf{t}',
\hat{\mathbf{t}}')\nn\\
 &= \res_z z^{-1}
X(\hat{\mathbf{t}};z)\tau(\mathbf{t},\hat{\mathbf{t}})X(\hat{\mathbf{t}}';-z)\tau(\mathbf{t}',
\hat{\mathbf{t}}'),
\end{align}
where $\mathbf{t}=(t_1, t_3, t_5, \cdots)$,
$\hat{\mathbf{t}}=(\hat{t}_1, \hat{t}_3, \hat{t}_5, \cdots)$, and
$X$ is a vertex operator given by
\[
X(\mathbf{t};z)=\exp\left(\sum_{k\in\Zop} t_k
z^k\right)\,\exp\left(-\sum_{k\in\Zop}\frac{2}{k\,z^k}\frac{\pd}{\pd
t_k}\right).
\]
Here the residue of a Laurent series is taken as
$\res_z(\sum_{i\in\Z}f_i z^i)=f_{-1}$. In \cite{Sh} Shiota proposed
a scalar Lax representation of the hierarchy~\eqref{2BKP-bltau},
though this did not attract much attention as it contains
pseudo-differential operators with derivations of two spatial
variables. Recently, a Lax pair representation of the two-component
BKP hierarchy was found by Liu, Zhang and one of the authors
\cite{LWZ}. It was shown that the hierarchy~\eqref{2BKP-bltau} can
be redefined by certain extension of the following Lax equations
(see Section~3 below):
\begin{align}\label{PPht0}
& \frac{\pd P}{\pd t_k}=[(P^k)_+, P], \quad \frac{\pd \hat{P}}{\pd t_k}=[(P^k)_+, \hat{P}],  \\
\label{PPhth0} & \frac{\pd P}{\pd \hat{t}_k}=[-(\hat{P}^k)_-, P],
\quad \frac{\pd \hat{P}}{\pd \hat{t}_k}=[-(\hat{P}^k)_-, \hat{P}]
\end{align}
with $k\in\Zop$, where
\begin{equation*}
P= D+\sum_{i\ge1}u_i  D^{-i}, \quad \hat{P}=
D^{-1}\hat{u}_{-1}+\sum_{i\ge1}\hat{u}_i D^i\hbox{ with }
D=\frac{\od}{\od x}
\end{equation*}
are pseudo-differential operators such that $P^*=- D P  D^{-1}$,
$\hat{P}^*=- D\hat{P} D^{-1}$. Note that the first equation in
\eqref{PPht0} is just the Lax formulation of the BKP hierarchy
appearing in \cite{DKJM-KPBKP}. Our arguments will be based on the
Lax pair representation \eqref{PPht0}, \eqref{PPhth0} of the
two-component BKP hierarchy.

Observe that the expression \eqref{PPht0}, \eqref{PPhth0} is similar
to the Lax pair representation of the two-dimensional Toda hierarchy
\cite{UT}, which carries a tri-Hamiltonian structure \cite{Carlet}.
Following the idea of \cite{Carlet}, we want to use the $R$-matrix
theory to construct Hamiltonian structures of the two-component BKP
hierarchy \eqref{PPht0}, \eqref{PPhth0}.

We are also motivated by the recent work \cite{CDM}, in which
Carlet, Dubrovin and Mertens constructed an infinite-dimensional
Frobenius manifold underlying the two-dimensional Toda hierarchy.
Due to the similarity of the Lax representations mentioned above, we
expect that there also exists an infinite dimensional Frobenius
manifold that underlies the two-component BKP hierarchy. A hint is
that the potential $F$ (in the notion of \cite{Ta}, namely the
dispersionless limit of the logarithm of the tau function, see
Section~3 below) of the dispersionless two-component BKP hierarchy
was discovered to satisfy certain infinite-dimensional WDVV-type
associativity equation \cite{CT}. While in the finite-dimensional
case, the concept of Frobenius manifolds \cite{Du} is known as a
geometric description of the WDVV equations, and associated to
certain nondegenerate Frobenius manifold there lies a Poisson pencil
so that a bihamiltonian hierarchy can be constructed \cite{DZ}. We
hope that this article and follow-up work might help to understand
the theory of infinite-dimensional manifolds.

This article is arranged as follows. In next section we recall the
definition and some properties of pseudo-differential operators
introduced in \cite{LWZ}, and in Section~3 we recall the Lax pair
representation of the two-component BKP hierarchy. In Sections 4 and
5, an $R$-matrix will be used to construct Poisson brackets on an
algebra of pseudo-differential operators, and then after appropriate
reductions of the Poisson brackets we obtain a bihamiltonian
structure of the two-component BKP hierarchy. In Section~6 we
compute the dispersionless limit of this bihamiltonian structure.
Finally some remarks are given in Section~7.

\section{Pseudo-differential operators}
For preparation we recall the notion of pseudo-differential
operators over a ring with certain gradation as introduced
in~\cite{LWZ}.

Let $\cA$ be a ring, and $D:\cA\to\cA$ be a derivation. The algebra
of usual pseudo-differential operators is
\begin{equation}\label{}
\cD^-=\left\{ \sum_{i<\infty} f_i D^i \mid f_i\in\cA\right\}.
\end{equation}
This algebra is topologically complete with a topological basis
given by the following filtration:
\[\cdots\subset\cD^-_{(d-1)}\subset\cD^-_{(d)}\subset\cD^-_{(d+1)}\subset\cdots,\quad
 \cD^-_{(d)}=\left\{\sum_{i \le d}f_i  D^i\mid f_i\in\cA\right\},
\]
and in this algebra two elements are multiplied as series of the
following product of monomials:
\begin{equation}\label{pro}
f D^i\cdot g D^j=\sum_{r\geq0}\binom{i}{r}f\, D^r(g)\, D^{i+j-r},
\quad f,g\in\cA.
\end{equation}

Assume there is a gradation on $\mathcal{A}$ such that
\[
\mathcal{A}=\prod_{i\geq0}\cA_i, \quad  D: \cA_i\to\cA_{i+1}, \quad
\cA_i\cdot\cA_j\subset\cA_{i+j},
\]
and consider the linear space
\[
\cD=\left\{\sum_{i\in\Z} f_i  D^i\mid f_i\in\cA\right\}.
\]
Obviously $\cD^-\subset\cD$.

For any $k\in\Z$, denote by $\cD_k$ the set of homogeneous operators
with degree $k$ in $\cD^-$, i.e.,
\[
\cD_k=\left\{\sum_{i\le k}f_i D^i\mid f_i\in\cA_{k-i}\right\}.
\]
Let $\cD^+$ be a subspace of $\cD$ that reads
\begin{equation}\label{Dhat}
\cD^+=\bigcup_{d\in\Z}\cD^+_{(d)},\quad \cD^+_{(d)}=\prod_{k \ge
d}\cD_k,
\end{equation}
and $\cD^+$ have a topological basis given by the filtration
\[\cdots\supset\cD^+_{(d-1)}\supset\cD^+_{(d)}\supset\cD^+_{(d+1)}\supset\cdots.
\]
In fact, every element $A\in\cD^+$ has the following normal
expansion \cite{LWZ}
\[
A=\sum_{i\in\Z}\left(\sum_{j\ge \max\{0,m-i\}}a_{i,j}\right)
D^i,\quad a_{i,j}\in\cA_j
\]
with some integer $m$. Note that $\cD_k\cdot\cD_l\subset\cD_{k+l}$
according to the multiplication defined by \eqref{pro}, then this
multiplication can be naturally extended to $\cD^+$ such that
$\cD^+$ becomes an associative algebra.

\begin{defn}[\cite{LWZ}]
Elements of $\cD^-$ (resp. $\cD^+$) are called pseudo-differential
operators of the first type (resp. the second type) over $\cA$. The
intersection of $\cD^-$ and $\cD^+$ in $\cD$ is denoted by
\[
\cD^b=\cD^-\cap \cD^+,
\]
and its elements are called  bounded pseudo-differential operators.

Sometimes to indicate the ring $\cA$ and the derivation $ D$, we
will use the notations $\cD^{\pm}(\cA,  D)$ instead of $\cD^\pm$.
\end{defn}

Pseudo-differential operators of the second type have similar
properties to those of the operators in $\cD^-$. For any operator
\begin{equation}
 A=\sum_{i\in\Z} f_i  D^i\in\cD^\pm,
\end{equation}
its positive part, negative part, residue and adjoint operator are
defined to be respectively
\begin{align}\label{Apm}
&A_+=\sum_{i\geq0} f_i  D^i, \quad A_-=\sum_{i<0} f_i  D^i, \\
&\res\,A=f_{-1},\quad A^*=\sum_{i\in\Z}(- D)^i\cdot f_i.
\end{align}
Note that the formulae \eqref{Apm} give two projections of $\cD$,
and they induce the following decompositions of spaces
\begin{equation}\label{Dpm}
\cD^\pm=(\cD^\pm)_+\oplus(\cD^\pm)_-.
\end{equation}
Particularly one sees that
\begin{equation}\label{subset} (\cD^-)_+\subset\cD^b,
\quad (\cD^+)_-\subset\cD^b.
\end{equation}

An element $A$ of $(\cD^\pm)_+$ is called a \emph{differential
operator}. Let $A(f)$ denote the action of a differential operator
$A$ on $f\in\cA$.

Elements of the quotient space $\cF=\cA/ (D(\cA)\oplus\C)$ are
called \emph{local functionals}, which are denoted as
\[\int f\, \od x=f+ D(\cA),\quad f\in\cA.\]

Introduce a map
\begin{equation}\label{int}
\la\, \,\ra:\ \cD\to\cF,\quad A\mapsto \la A\ra=\int \res A\, \od x.
\end{equation}
Then the pairing
\begin{equation}
\la A,B\ra=\la A B\ra\label{inpro}
\end{equation}
defines an inner product on each of $\cD^\pm$.

Given any subspace $\mathcal{S}\subset\cD^\pm$, we denote by
$\mathcal{S}^*$ the dual space of $\mathcal{S}$ (c.f. the notation
of adjoint operators). Via the above inner product, we have the
following identification of dual spaces
\begin{equation}\label{Ddual}
(\cD^\pm)^*=\cD^\pm.
\end{equation}
Consider the decompositions \eqref{Dpm}, it is easy to see that
\[
\big((\cD^\pm)_\pm\big)^*=(\cD^\pm)_\mp.
\]

We also decompose $\cD^\pm$ as
\begin{equation}\label{Deo}
\cD^\pm=\cD^\pm_0\oplus\cD^\pm_1,
\end{equation}
where
\[
\cD^\pm_\nu=\left\{ A\in\cD^\pm\mid A^*=(-1)^\nu A\right\}, \quad
\nu=0,1.
\]
Since $\la A\ra=-\la A^*\ra$ for any $A\in\cD^\pm$, then the dual
subspaces of $\cD^\pm_\nu$ read
\begin{equation}\label{}
(\cD^\pm_\nu)^*=\cD^\pm_{1-\nu}, \quad \nu=0,1.
\end{equation}

For more details on properties of pseudo-differential operators one
can refer to~\cite{Dickey, LWZ}.

\section{The two-component BKP hierarchy}

The two types of pseudo-differential operators serve in \cite{LWZ}
to give a scalar Lax pair representation of the two-component BKP
hierarchy, which is reviewed as follows.

Let $\tilde{M}$ be an infinite-dimensional manifold with local
coordinates
\[(a_1, a_3, a_5, \dots, b_1, b_3, b_5, \dots),\]
and $\tilde{\cA}$ be the algebra of differential polynomials on
$\tilde{M}$:
\[\tilde{\cA}=C^{\infty}(\tilde{M})[[a_{k}^{(s)}, b_{k}^{(s)}\mid k\in\Zop, s\ge1]].\]
We assign a gradation on $\tilde{\cA}$ by
\[
\deg f=0 \hbox{ for } f\in C^{\infty}(\tilde{M}), \quad \deg
a_{k}^{(s)}=\deg b_{k}^{(s)}=s
\]
which make $\tilde{\cA}$ a topologically complete algebra:
\[
\tilde{\cA}=\prod_{i\ge0}\tilde{\cA}_i,\quad
\tilde{\cA}_i\cdot\tilde{\cA}_j\subset\tilde{\cA}_{i+j}.
\]
Note that this gradation is induced from the derivation
\[
 D: \tilde{\cA}\to\tilde{\cA}, \quad
 D=\sum_{s\ge0}\sum_{k\in\Zop} \left(a_k^{(s+1)}\frac{\pd}{\pd
a_k^{(s)}}+b_k^{(s+1)}\frac{\pd}{\pd b_k^{(s)}}\right)
\]
with $a_k^{(0)}=a_k$, $b_k^{(0)}=b_k$. So one can define the
algebras $\tilde{\cD}^\pm=\cD^\pm(\tilde{\cA}, D)$ of
pseudo-differential operators as was done in last section.

Introduce two operators
\begin{align}
\Phi=1+\sum_{i\ge 1}a_i D^{-i} \in\tilde{\cD}^-,\quad
\Psi=1+\sum_{i\ge 1}b_i D^{i} \in\tilde{\cD}^+,
\end{align}
where $a_2, a_4, a_6, \dots, b_2, b_4, b_6, \dots\in \tilde{\cA}$
are determined by the following conditions
\begin{equation}
\Phi^*= D\Phi^{-1} D^{-1},\quad \Psi^*= D\Psi^{-1} D^{-1}.
\label{phipsi}
\end{equation}
Then the two-component BKP hierarchy \eqref{2BKP-bltau} can be
redefined to be
\begin{align}
&\frac{\pd \Phi}{\pd t_k}=- (P^k)_-\Phi, \quad \frac{\pd \Psi}{\pd t_k}=\bigl((P^k)_+ -\dt_{k1} \hat{P}^{-1}\bigr)\Psi, \label{ppt1}\\
&\frac{\pd \Phi}{\pd \hat{t}_k}=- (\hat{P}^k)_-\Phi, \quad \frac{\pd
\Psi}{\pd \hat{t}_k}=(\hat{P}^k)_+\Psi, \label{ppt2}
\end{align}
where $k\in\Zop$, and the operators $P$, $\hat{P}$ read
\begin{equation} \label{PPh}
P=\Phi D\Phi^{-1} \in \tilde{\cD}^-,\quad \hat{P}=\Psi
D^{-1}\Psi^{-1} \in \tilde{\cD}^+.
\end{equation}

The operators $P$, $\hat{P}$ have the following expressions:
\begin{equation}
P= D+\sum_{i\ge1}u_i  D^{-i}, \quad \hat{P}=
D^{-1}\hat{u}_{-1}+\sum_{i\ge1}\hat{u}_i D^i,
\end{equation}
with $\hat{u}_{-1}=(\Psi^{-1})^*(1)$, and they satisfy
\begin{equation}\label{PQstar}
P^*=- D P  D^{-1}, \quad \hat{P}^*=- D\hat{P} D^{-1},
\end{equation}
which implies
\begin{equation}\label{PkQk}
(P^k)_+(1)=0,\quad (\hat{P}^k)_+(1)=0, \quad k\in\Zop.
\end{equation}

Observe that the coefficients of $P$ and $\hat{P}$ are elements of
the algebra $\tilde{\cA}$, and among these coefficients the ones
with odd subscript are independent, while the others are determined
by the conditions \eqref{PQstar}. Assume that
\begin{equation}\label{equ}
\mathbf{u}=(u_1, u_3, \dots, \hat{u}_{-1}, \hat{u}_1, \hat{u}_3,
\dots)
\end{equation}
serves as a coordinate of some infinite-dimensional manifold $M$,
then the algebra $\cA$ of differential polynomials on $M$ reads
\[
\cA=C^\infty(M)[[\mathbf{u}^{(s)}\mid s\ge1]],
\]
which is a subalgebra of $\tilde{\cA}$. Similarly as above, one can
assign a gradation to $\cA$ that is induced from the derivation
\[
D: \cA\to\cA,\quad
D=\sum_{s\ge0}\mathbf{u}^{(s+1)}\cdot\frac{\pd}{\pd\mathbf{u}^{(s)}}
\]
with $\mathbf{u}^{(0)}=\mathbf{u}$, and then define the algebras
$\cD^\pm=\cD^\pm(\cA, D)$ of pseudo-differential operators over
$\cA$.

Clearly $P\in\cD^-$, $\hat{P}\in\cD^+$. When the two-component BKP
hierarchy \eqref{ppt1}, \eqref{ppt2} is restricted from
$\tilde{\cA}$ to $\cA$, it becomes
\begin{align}\label{PPht}
& \frac{\pd P}{\pd t_k}=[(P^k)_+, P], \quad \frac{\pd \hat{P}}{\pd t_k}=[(P^k)_+, \hat{P}],  \\
\label{PPhth} & \frac{\pd P}{\pd \hat{t}_k}=[-(\hat{P}^k)_-, P],
\quad \frac{\pd \hat{P}}{\pd \hat{t}_k}=[-(\hat{P}^k)_-, \hat{P}]
\end{align}
with $k\in\Zop$. In the present article we regard the two-component
BKP hierarchy as the evolutionary equations \eqref{PPht},
\eqref{PPhth} defined on the algebra $\cA$.

In fact, the hierarchy \eqref{PPht}, \eqref{PPhth} possesses a tau
function $\tau=\tau(\mathbf{t},\hat{\mathbf{t}})$ defined by
\begin{align}\label{tau}
\om=\od(2\,\pd_x\log\tau) ~~ \hbox{   with  } ~~ x=t_1,
\end{align}
where $\om$ is the following closed $1$-form:
\[
\om=\sum_{k\in\Zop}(\res\, P^k\, \od t_k+\res\, \hat{P}^k\,
\od\hat{t}_k).
\]
This tau function solves the bilinear equation \eqref{2BKP-bltau},
which is the original definition of the two-component BKP hierarchy.

\begin{rmk}
The dispersionless limit of the flows \eqref{PPht}, \eqref{PPhth}
first exists in \cite{Ta}, where Takasiki also considered the
dispersionless limit of the logarithm of the tau function as given
in \eqref{tau}. Inspired by \cite{Ta}, Chen and Tu \cite{CT}
discovered that the leading term of $\log\tau$ solves an
infinite-dimensional associativity equation of WDVV type.
\end{rmk}

\section{$R$-matrix and pseudo-differential operators}
To show that the two-component BKP hierarchy \eqref{PPht},
\eqref{PPhth} possesses a bihamiltonian structure, we need to
construct a Poisson pencil for it. The method is to use the standard
$R$-matrix theory and introduce Poisson brackets on a Lie algebra
(see \cite{STS, LP, OR} and references therein), then restrict the
Poisson brackets to certain submanifold of the Lie algebra. Our
approach is similar with that used by Carlet \cite{Carlet} for the
two-dimensional Toda hierarchy.

We first recall the $R$-matrix formalism. Let $\fg$ be a Lie
algebra, and $R:\fg\to\fg$ be a linear transformation. Then
 $R$ is called an $R$-matrix~\cite{STS} on $\fg$ if it defines a Lie
bracket by
\begin{equation}\label{}
[X,Y]_R=[R(X), Y]+[X, R(Y)],\quad X, Y\in\fg.
\end{equation}
A sufficient condition for a transformation $R$ being an $R$-matrix
is that $R$ solves the modified Yang-Baxter equation \cite{STS}
\begin{equation}\label{YBeq}
[R(X), R(Y)]-R([X, Y]_R)=-[X,Y]
\end{equation}
for all $X$, $Y\in\fg$.

Assume that $\fg$ is an associative algebra, with the Lie bracket
defined naturally by commutators, and there is a map
$\la~\ra:\fg\to\C$ that defines a non-degenerate symmetric invariant
bilinear form (inner product) $\la\,,\ra$ by
\[
\langle X, Y\rangle=\la X Y\ra=\la Y X\ra,\quad X, Y\in\fg.
\]
Via this inner product one can identify $\fg$ with its dual space
$\fg^*$. The tangent and the cotangent bundles of $\fg$ are denoted
by $T\fg$ and $T^*\fg$ respectively, with fibers $T_A\fg=\fg$ and
$T^*_A\fg=\fg^*$ at every point $A\in\fg$.

Let $R^*$ be the adjoint transformation of $R$ with respect to the
above inner product. We introduce the notations of the symmetric and
the anti-symmetric parts of $R$ respectively as
\[ R_s=\frac1{2}(R+R^*), \quad R_a=\frac1{2}(R-R^*). \]

The $R$-matrix formalism is briefly stated as follows. Given an
$R$-matrix $R:\fg\to\fg$ that satisfies certain conditions, there
define three compatible Poisson brackets on $\fg$, say, the linear,
the quadratic and the cubic brackets in the notion of \cite{LP, OR}.

In particular, let us recall the quadratic bracket, which will be
used to construct a Poisson pencil for the two-component BKP
hierarchy.
\begin{lem}[\cite{LP, OR}]\label{thm-qbr}
Let $f$, $g$ be two arbitrary smooth functions on $\fg$, and $\nabla
f, \nabla g \in T_A^*\fg$ be their gradients at any point $A\in\fg$.
Given a linear transformation $R:\fg\to\fg$, if both $R$ and its
anti-symmetric part $R_a$ satisfy the modified Yang-Baxter
equation~\eqref{YBeq}, then the quadratic bracket
  \begin{equation}\label{quabr}
\{f, g\}(A)=\frac1{4}\big(\la[A,\nabla f],R(A\nabla g+\nabla g\cdot
A)\ra-\la[A,\nabla g],R(A\nabla f+\nabla f\cdot A)\ra\big)
\end{equation}
defines a Poisson bracket on $\fg$.
\end{lem}
Note that the bracket~\eqref{quabr} can be rewritten as
\begin{equation*}
\{f, g\}(A)=\la \nabla f,\cP_A(\nabla g)\ra,
\end{equation*}
where $\cP:T^*\fg\to T\fg$ is a Poisson tensor given by
\begin{align*}\label{}
\cP_A(\nabla g)=&-\frac1{4}[A, R(A\nabla g+\nabla g\cdot
A)]-\frac1{4}A R^*([A,\nabla g])-\frac1{4}R^*([A,\nabla g]) A,
\end{align*}
namely,
\begin{align}\label{Poiten}
\cP_A(\nabla g)=&-\frac1{2}A\big(R_s(A\nabla g)+ R_a(\nabla g\cdot
A)\big)+\frac1{2} \big(R_a(A\nabla g)+ R_s(\nabla g\cdot A)\big)A.
\end{align}

Henceforth we take $\fg$ to be the algebra
\[
\fD=\cD^-\times\cD^+,
\]
where $\cD^-$ and $\cD^+$ are the sets of pseudo-differential
operators of the first type and the second type over some
differential algebra $\cA$ as defined in Section~2. In $\fD$ the
elements read $\bX=(X,\hat{X})$, and the operations are defined
diagonally as
\[
(X,\hat{X})+(Y,\hat{Y})=(X+Y,\hat{X}+\hat{Y}),\quad
(X,\hat{X})(Y,\hat{Y})=(X Y,\hat{X} \hat{Y}).
\]
So $\fD$ is indeed an associative algebra. Moreover, the algebra
$\fD$ is equipped with an inner product define by
\[
\la(X,\hat{X}),(Y,\hat{Y})\ra=\la(X,\hat{X})(Y,\hat{Y})\ra=\la X,
Y\ra+\la\hat{X},\hat{Y}\ra,
\]
see \eqref{int}, \eqref{inpro}. Via this inner product we have the
identification of dual spaces as above:
\[
\fD^*=(\cD^-)^*\times(\cD^+)^*=\cD^-\times\cD^+=\fD.
\]

Inspired by \cite{Carlet}, we introduce a linear transformation of
$\fD$ as follows
\begin{equation}\label{Rmatrix}
R:\fD\to\fD, \quad (X,\hat{X})\mapsto(X_+ -X_- +2\hat{X}_-,\hat{X}_-
-\hat{X}_+ +2 X_+).
\end{equation}
Since $R=\Pi-\tilde{\Pi}$, where
\[
\Pi(X,\hat{X})=(X_+ +\hat{X}_-,\hat{X}_- +X_+),\quad
\tilde{\Pi}(X,\hat{X})=(X_- -\hat{X}_-,\hat{X}_+- X_+)
\]
are two projections of $\fD$ onto its subalgebras, more exactly,
\begin{align*}\label{}
&\Pi\fD=\{(X,X)\mid X\in\cD^b\}, \quad
\tilde{\Pi}\fD=(\cD^-)_-\times(\cD^+)_+,
\\
&\Pi^2=\Pi, ~~\tilde{\Pi}^2=\tilde{\Pi},
~~\tilde{\Pi}\,\Pi=0=\Pi\,\tilde{\Pi},
~~\Pi+\tilde{\Pi}=\mathrm{id},
\end{align*}
then transformation $R$ satisfies the modified Yang-Baxter
equation~\eqref{YBeq}. Hence $R$ is an $R$-matrix on~$\fD$.

On the other hand, with respect to the inner product on $\fD$ the
adjoint transformation of $R$ reads
\[
R^*:\fD\to\fD, \quad (X,\hat{X})\mapsto(X_- -X_+
+2\hat{X}_-,\hat{X}_+ -\hat{X}_- +2 X_+).
\]
Then the symmetric and anti-symmetric parts of $R$ are given by
\begin{equation}\label{RaRs}
R_s(X,\hat{X})=2(\hat{X}_-,X_+), \quad
R_a(X,\hat{X})=(X_+-X_-,\hat{X}_--\hat{X}_+).
\end{equation}
Observe that $R_a$ can be expressed as the difference of two
projections onto subalgebras of $\fD$, hence $R_a$ also solves the
Yang-Baxter equation~\eqref{YBeq}. Thus the $R$-matrix given in
\eqref{Rmatrix} fulfills the condition of Lemma~\ref{thm-qbr}.

We regard $\fD$ as an infinite-dimensional manifold, whose
coordinate is given by the coefficients of the general expression of
its elements
\begin{equation}\label{eqA}
\bA=\left(\sum_{i\in\Z}w_i D^i,\, \sum_{i\in\Z}\hat{w}_i
D^i\right)\in\fD.
\end{equation}
The set $\cF$ of local functionals over the differential algebra
$\cA$ (see Section~2) plays the role of $C^\infty(\fg)$. For any
$F=\int f\,\od x\in\cF$, the variational gradient of $F$ at $\bA$
given in \eqref{eqA} is defined to be
\[
\frac{\dt F}{\dt\bA}=\left(\sum_{i\in\Z} D^{-i-1}\frac{\dt F}{\dt
w_i(x)},\, \sum_{i\in\Z}D^{-i-1}\frac{\dt
F}{\dt\hat{w}_i(x)}\right),
\]
where $\dt F/\dt w(x)=\sum_{j\ge0}(-D)^j\left(\pd f/\pd
w^{(j)}\right)$. Note that $\dt F/\dt \bA$ is not contained in
$\fD^*=\fD$ in general, so to go forward we need to do some
restriction.

It shall be indicated that, in this paper we only consider
functionals with variational gradients lying in $\fD$. Let $\cF_0$
denote the set of such functionals.

Now we can use Lemma~\ref{thm-qbr} and the formulae \eqref{Poiten},
\eqref{RaRs} to obtain the following result.
\begin{lem}\label{thm-qbrD}
Let $F$ and $G$ be two arbitrary functionals in $\cF_0$. On the
algebra $\fD$ there is a quadratic Poisson bracket
\begin{equation}\label{Poibra}
\{F, G\}(\bA)=\left\la\frac{\dt F}{\dt\bA}, \cP_\bA\left(\frac{\dt
G}{\dt\bA}\right)\right\ra, \quad \bA=(A,\hat{A})\in\fD,
\end{equation}
 where the Poisson
tensor $\cP: T\fD^*\to T\fD$ is defined by
\begin{align}\label{PAX}
\cP_{(A,\hat{A})}(X,\hat{X})=&
\big(A(X A)_--(A X)_-A-A(\hat{A}\hat{X})_-+(\hat{X}\hat{A})_-A, \nn\\
&~~\hat{A}(\hat{X}\hat{A})_+-(\hat{A}\hat{X})_+\hat{A}-\hat{A}(A
X)_++(X A)_+\hat{A}\big).
\end{align}
\end{lem}

Aiming at Hamiltonian structures of the two-component BKP hierarchy,
we need to reduce the Poisson structure \eqref{PAX} to an
appropriate submanifold of $\fD$. Recall the decompositions
\eqref{Deo}, let us decompose the space $\fD$ as
\begin{equation}\label{bddec}
\fD=\fD_0\oplus\fD_1,
\end{equation}
where $\fD_\nu=\cD^-_\nu\times\cD^+_{\nu}$ for $\nu=0,1$. Since the
subspaces $\fD_0$ and $\fD_1$ are dual to each other with respect to
the inner product on $\fD$, then for any $\bA\in\fD_\nu$ we have
$T_\bA^*\fD_\nu=(\fD_\nu)^*=\fD_{1-\nu}$ for $\nu=0,1$. It is
straightforward to verify the following lemma.
\begin{lem}\label{thm-qbrD01}
  The Poisson structure \eqref{PAX} on $\fD$ can be
  properly restricted to each of its subspaces $\fD_0$ and $\fD_1$.
\end{lem}

\section{Bihamiltonian representation of the two-component BKP hierarchy}
In this section, we are to find a submanifold of $\fD$ where the
Poisson pencil for the two-component BKP hierarchy lies, then after
a further reduction of the Poisson structure constructed in last
section we will express the hierarchy \eqref{PPht}, \eqref{PPhth} to
the form of Hamiltonian equations.

Recall the operators $P\in\cD^-$, $\hat{P}\in\cD^+$ given in
\eqref{PPh}, we let
\begin{equation}\label{eqbK}
\bA=(P^2 D^{-1}, D\hat{P}^2).
\end{equation}
It is easy to see that $\bA\in\fD_1$ (see \eqref{bddec}), and
$\bA=(A,\hat{A})$ has the following expression:
\begin{align}\label{eqK}
&A=P^2 D^{-1}= D+\sum_{i\ge0}(v_{-i} D^{-2i-1}+f_{-i} D^{-2i-2}),
\\
&\hat{A}=D\hat{P}^2=\rho D^{-1}\rho+\sum_{i\ge1}(\hat{v}_{i}
D^{2i-1}+\hat{f}_{i} D^{2i-2}), \quad \rho=\hat{u}_{-1}.
\label{eqKh}
\end{align}
Denote $\mathbf{v}=(v_{0}, v_{-1}, \dots, \hat{v}_0,
\hat{v}_1,\dots)$ with $\hat{v}_{0}=\rho^2$. Then the coordinate
$\mathbf{v}$ is related to $\mathbf{u}$ given in \eqref{equ} by a
Miura-type transformation, while $f_{-i}$ and $\hat{f}_i$ are linear
functions of derivatives of $\mathbf{v}$ determined by the symmetry
property $(A^*,\hat{A}^*)=-(A,\hat{A})$. Hence the flows of the
hierarchy \eqref{PPht}, \eqref{PPhth} can be described in the
coordinate $\mathbf{v}$.

Given any local functional $F\in\cF_0$ (remind the notation $\cF_0$
in last section), its variational gradient with respect to $\bA$,
say $\dt F/\dt\bA$, is defined to be $\bX=(X,\hat{X})\in\fD$ with
\begin{align}\label{vgX}
X=&\frac1{2}\sum_{i\ge0}\left(\frac{\dt F}{\dt
v_{-i}(x)} D^{2i}+ D^{2i}\frac{\dt F}{\dt v_{-i}(x)}\right), \\
\hat{X}=&\frac1{2}\sum_{i\ge0}\left(\frac{\dt F}{\dt \hat{v}_{i}(x)}
D^{-2i}+ D^{-2i}\frac{\dt F}{\dt \hat{v}_{i}(x)}\right).
\label{vgXh}
\end{align}
In a coordinate-free way, $\dt F/\dt\bA=\bX$ can be defined by
\begin{align}\label{}
\dt F=\la\bX,\dt\bA\ra, \quad \bX\in \fD_0.
\end{align}
Note that in the latter definition, the variational gradient is
determined up to a kernel part $\mathbf{Z}=(Z,\hat{Z})\in\fD_0$ such
that
\begin{equation}\label{vgker}
 Z_+=0, \quad \hat{Z}_-=0,\quad \hat{Z}_+(\rho)=0.
\end{equation}

Let us consider the coset $(D,0)+\cU$ consisting of operators of the
form \eqref{eqbK}, namely,
\begin{equation}\label{eqU}
\cU=(\cD^-_1)_-\times\big((\cD^+_1)_+\times \cM \big), \quad
\cM=\{\rho D^{-1}\rho\mid\rho\in\cA\}.
\end{equation}
Here $\cM$ is regarded as a $1$-dimensional manifold with coordinate
$\rho$, and this manifold has tangent spaces of the form
\[
T_\rho \cM=\{\rho D^{-1}f+f D^{-1}\rho\mid f\in\cA\}.
\]
So the tangent bundle, denoted by $T\cU$, of the coset $(D,0)+\cU$
has fibers
\begin{equation}\label{tanU}
T_{\bA}\cU=(\cD^-_1)_-\times\big((\cD^+_1)_+\times T_\rho \cM \big),
\quad \bA\in (D,0)+\cU,
\end{equation}
while the cotangent bundle $T^*\cU$ of $(D,0)+\cU$ is composed of
\begin{equation}\label{ctanU}
T_{\bA}^*\cU=(\cD^-_0)_+\times\big((\cD^+_0)_-\times T_\rho^*\cM
\big), \quad T^*_\rho\cM=\cA.
\end{equation}
From \eqref{vgX}, \eqref{vgXh} one sees that $\dt F/\dt\bA\in
T_{\bA}^*\cU$ for any $F\in\cF_0$.

Now we are ready to do the desired reduction of the Poisson
structure.
\begin{lem}\label{lem-Poicoset}
The map
\begin{equation}\label{Pred}
\cP: T^*\cU\to T\cU
\end{equation}
defined by the formula \eqref{PAX} is a Poisson tensor on the coset
 $( D,0)+\cU$ that consists of operators of the form
 \eqref{eqbK}.
\end{lem}
\begin{prf}
We only need to show that the map defined by \eqref{PAX} admits the
restriction to the coset $(D,0)+\cU$, i.e., the following map is
well defined:
\begin{equation}\label{PredA}
\cP_\bA: T^*_\bA\cU\to T_\bA\cU, \quad \bA\in(D,0)+\cU.
\end{equation}

Assume $\bX=(X,\hat{X})\in T^*_\bA\cU\subset\fD_0$. It follows from
Lemma~\ref{thm-qbrD01} that $\cP_\bA(X)\in\fD_1$. More precisely,
the first component of $\cP_\bA(X)$ belongs to $(\cD^-_1)_-$. On the
other hand, for any $\hat{Y}\in(\cD^+)_+$ we have
\begin{align*}\label{}
(\hat{A}\hat{Y}+\hat{Y}^*\hat{A})_-=&(\rho D^{-1}\rho\hat{Y}+\hat{Y}^*\rho D^{-1}\rho)_-\\
=&-(\hat{Y}^*\rho D^{-1}\rho)^*_-+\hat{Y}^*(\rho) D^{-1}\rho \\
 =&\rho D^{-1}\hat{Y}^*(\rho)+\hat{Y}^*(\rho) D^{-1}\rho\in
T_\rho \cM,
\end{align*}
then by taking $\hat{Y}=(\hat{X}\hat{A})_+, (A X)_+$ it follows that
the second component of $\cP_{\bA}(\bX)$ lies in $(\cD^+_1)_+\times
T_\rho \cM $. Thus $\cP_{\bA}(\bX)\in T_\bA\cU$, i.e., the map
\eqref{PredA} is well defined. The lemma is proved.
\end{prf}

\begin{rmk}
The proof of this lemma is the simplest case of the Dirac reduction
procedure for Poisson tensors, see e.g. \cite{OR}. In fact, one can
express the manifolds $\fD_1$ and $\fD_1^*$ as
\begin{align}\label{D1D1star}
\fD_1=\cU\times\mathcal{V}=T_{\bA}\cU\times \mathcal{V}_\bA,\quad
\fD_1^*=\fD_0= T_{\bA}^*\cU\times\cV^*_\bA,
\end{align}
where
\begin{align*}\label{}
&\cV=\cV_\bA=(\cD^-_1)_+\times \cN, \quad
\cN=\{X\in(\cD^+_1)_-\mid\res X=0\}, \\
& \cV^*_\bA=(\cD^-_0)_-\times(T_\rho^*)^\bot \cM, \quad
(T_\rho^*)^\bot \cM=\{\hat{Y}\in(\cD^+_0)_+\mid \hat{Y}(\rho)=0\}.
\end{align*}
Similar as the proof of Lemma~\ref{lem-Poicoset}, one can show that
the map
\begin{equation*}\label{}
\cP_{\bA}=\left(
              \begin{array}{cc}
                \cP_{\bA}^{\cU\cU} & \cP_{\bA}^{\cU\mathcal{V}} \\
                \cP_{\bA}^{\mathcal{V}\cU} & \cP_{\bA}^{\mathcal{V}\mathcal{V}} \\
              \end{array}
            \right): T_{\bA}^*\cU\times \cV^*_\bA\to
            T_{\bA}\cU\times\cV_\bA
\end{equation*}
defined by \eqref{PAX} is diagonal. Hence from Lemma
\ref{thm-qbrD01} it follows that the map \eqref{PAX} gives a Poisson
tensor on the coset $(D,0)+\cU\subset\cD_1$.
\end{rmk}

\begin{lem}
On the coset $(D,0)+\cU$ there are two compatible Poisson tensors
defined by the following formulae:
\begin{align}\label{Poi1}
\cP_1(X,\hat{X})=&
\big(A(X D^{-1})_-+ D^{-1}(X A)_- -( D^{-1} X)_-A- (A X)_- D^{-1} \nn\\
&~~-A( D\hat{X})_-- D^{-1}(\hat{A}\hat{X})_-+(\hat{X} D)_-A+(\hat{X}\hat{A})_- D^{-1}, \nn\\
&~~\hat{A}(\hat{X} D)_++ D(\hat{X}\hat{A})_+-( D\hat{X})_+\hat{A}-(\hat{A}\hat{X})_+ D \nn\\
&~~-\hat{A}( D^{-1}X)_+- D(A X)_++(X D^{-1})_+\hat{A}+(X A)_+ D\big), \\
\cP_2(X,\hat{X})=&
\big(A(X A)_--(A X)_-A-A(\hat{A}\hat{X})_-+(\hat{X}\hat{A})_-A, \nn\\
&~~\hat{A}(\hat{X}\hat{A})_+-(\hat{A}\hat{X})_+\hat{A}-\hat{A}(A
X)_++(X A)_+\hat{A}\big) \label{Poi2}
\end{align}
with $(X,\hat{X})\in T^*_\bA\cU$ at any point
$\bA=(A,\hat{A})\in(D,0)+\cU$.
\end{lem}
\begin{prf}
Lemma~\ref{lem-Poicoset} shows that $\cP_2$ is a Poisson tensor on
the coset $( D,0)+\cU$.

Introduce a shift transformation on $( D,0)+\cU$ as
\[
\mathscr{S}: (A,\hat{A})\mapsto (A+s D^{-1},\hat{A}+s D)
\]
with $s$ being a parameter. Then the push-forward of the Poisson
tensor $\cP_2$ reads
\begin{equation}\label{Poishift}
(\mathscr{S}_*\cP_2)(X,\hat{X})=\cP_2(X,\hat{X})+s
\cP_1(X,\hat{X})+s^2 \cP_0(X,\hat{X}),
\end{equation}
where
\begin{align*}\label{}
\cP_0(X,\hat{X})=&
\big( D^{-1}(X  D^{-1})_--( D^{-1} X)_- D^{-1}- D^{-1}( D \hat{X})_-+(\hat{X} D )_- D^{-1}, \nn\\
&~~ D (\hat{X} D )_+-( D \hat{X})_+ D - D ( D^{-1} X)_++(X
 D^{-1})_+ D \big).
\end{align*}
By virtue of the symmetry property $(X^*,\hat{X}^*)=(X,\hat{X})$
that yields the formulae
\begin{align*}\label{}
&(X D^{-1})_\pm=X_\pm D^{-1}\mp X_+(1) D^{-1}, \\
& ( D^{-1}X)_\pm= D^{-1}X_\pm\mp D^{-1}\cdot X_+(1), \\
&( D\hat{X})_\pm= D\hat{X}_\pm, \quad (\hat{X} D)_\pm=\hat{X}_\pm D,
\end{align*}
one can check $\cP_0(X,\hat{X})=0$. Hence the expansion
\eqref{Poishift} implies that $\cP_1$ is a Poisson tensor that is
compatible with $\cP_2$. The lemma is proved.
\end{prf}

Let $\{\cdot, \cdot\}_{1,2}$ denote the Poisson brackets given in
\eqref{Poibra} with Poisson tensors being $\cP_{1,2}$ respectively.
We arrive at the main result of this article.
\begin{thm}\label{thm-Poi2BKP}
The two-component BKP hierarchy \eqref{PPht}, \eqref{PPhth} can be
expressed in the following bihamiltonian recursion form
\begin{align}\label{Ham1}
\frac{\pd F}{\pd t_k}=\{F, H_{k+2}\}_1(\bA)=\{F, H_{k}\}_2(\bA),
\\
\frac{\pd F}{\pd \hat{t}_k}=\{F, \hat{H}_{k+2}\}_1(\bA)=\{F,
\hat{H}_{k}\}_2(\bA) \label{Ham2}
\end{align}
with $k\in\Zop$, where $F\in\cF_0$, $\bA=(P^2 D^{-1}, D\hat{P}^2)$
as given in \eqref{eqbK}, and the Hamiltonians are
\begin{align}\label{HHhat}
H_{k}=\frac{2}{k}\la P^k\ra, ~~\hat{H}_{k}=-\frac{2}{k}\la
\hat{P}^k\ra, \quad k\in\Zop.
\end{align}
\end{thm}
\begin{prf}
First let us compute the variational gradients of the Hamiltonian
functionals. Since
\[
\dt H_{k}=\la P^{k-2},\dt P^2\ra=\la D P^{k-2},\dt (P^2
D^{-1})\ra=\la (D P^{k-2},0),\dt\bA\ra
\]
and similarly
\[
\dt\hat{H}_{k}= \la(0, -\hat{P}^{k-2} D^{-1}), \dt\bA\ra,
\]
then up to kernel parts given in \eqref{vgker} we have the
variational gradients of the Hamiltonians:
\begin{equation}\label{dtHHhat}
\frac{\dt H_{k}}{\dt\bA}=(D P^{k-2},0),\quad
\frac{\dt\hat{H}_{k}}{\dt\bA}= (0, -\hat{P}^{k-2} D^{-1})
\end{equation}
One can easily see that different choices of the kernel parts do not
change the definition of the Poisson tensors $\cP_{1,2}$.

According to the flows \eqref{PPht}, \eqref{PPhth} one has
\[
\frac{\pd\bA}{\pd t_k}=\left([(P^k)_+,P^2]D^{-1},
D[(P^k)_+,\hat{P}^2]\right).
\]
Note that
\[
\frac{\pd F}{\pd t_k}=\left\la \frac{\dt
F}{\dt\bA},\frac{\pd\bA}{\pd t_k}\right\ra,
\]
then to show \eqref{Ham1} we only need to verify the equations
\begin{align}\label{Hflow}
&\frac{\pd\bA}{\pd t_k}=\cP_1\left(\frac{\dt
H_{k+2}}{\dt\bA}\right)=\cP_2\left(\frac{\dt H_{k}}{\dt\bA}\right).
\end{align}
The verification is straightforward by substituting~\eqref{dtHHhat}
into~\eqref{Poi1}, \eqref{Poi2} with the help of the following
formulae induced from \eqref{PkQk}:
\[
( D P^k D^{-1})_\pm= D (P^k)_\pm D^{-1}, \quad ( D \hat{P}^k
D^{-1})_\pm= D (\hat{P}^k)_\pm D^{-1}, \quad k\in\Zop.
\]
The equations \eqref{Ham2} can be checked similarly. The theorem is
proved.
\end{prf}

This theorem implies that the tau function \eqref{tau} of the
two-component BKP hierarchy is defined from the tau-symmetry of
Hamiltonian densities \cite{DZ} (up to the signs of $\hat{H}_k$).

\begin{rmk}
One can also construct Hamiltonian structures of the two-component
BKP hierarchy by reducing the linear and the cubic Poisson brackets
induced from the $R$-matrix mentioned in last section. However, from
these brackets we have not found bihamiltonian recursion relations
like \eqref{Ham1}, \eqref{Ham2}.
\end{rmk}

\section{Dispersionless limit of the bihamiltonian structure}

Let us compute the leading term of the bihamiltonian structure in
\eqref{Ham1}, \eqref{Ham2} of the two-component BKP hierarchy, which
would make sense in studying the corresponding Frobenius manifold if
there be.

First we replace the pseudo-differential operators by Laurent series
of symbols. In the dispersionless case, the operator
$\bA=(P^2D^{-1},D\hat{P}^2)$ becomes
\begin{align}\label{}
(a(z),\hat{a}(z))=\left(z+\sum_{i\ge0} v_{-i}z^{-2i-1},
\sum_{i\ge0}\hat{v}_{i}z^{2i-1}\right),
\end{align}
and the coordinate-type local functionals $v_{-i}(y)$,
$\hat{v}_j(y)$ have variational gradients $(z^{2i}\dt(x-y),0)$, $(0,
z^{-2j}\dt(x-y))$ respectively. Substituting these Laurent series
into the Poisson brackets defined by the formulae \eqref{Poibra},
\eqref{Poi1}, \eqref{Poi2}, we obtain the following result.

For the convenience of expression we set $v_1=1$, $v_i=0$ when
$i\geq2$, and $\hat{v}_j=0$ when $j\leq-1$.
\begin{itemize}
\item[i)] The first bracket: for $i,j\geq0$,
\begin{align}\label{}
&\{v_{-i}(x),v_{-j}(y)\}_1^{[0]}=(1-\dt_{i0}-\dt_{j0})\big(2(i+j-1)v_{-i-j+1}(x)\,\dt'(x-y) \nn\\
&\qquad\qquad\qquad\qquad+(2j-1)v_{-i-j+1}'(x)\,\dt(x-y)\big),\\
&\{\hat{v}_{i}(x),\hat{v}_{j}(y)\}_1^{[0]}=-(1-\dt_{i0}-\dt_{j0})\big(2(i+j-1)\hat{v}_{i+j}(x)\,\dt'(x-y)
\nn\\
&\qquad\qquad\qquad\qquad+(2j-1)\hat{v}_{i+j}'(x)\,\dt(x-y)\big),\\
&\{v_{-i}(x),\hat{v}_{j}(y)\}_1^{[0]}\nn\\
&\qquad=2(i-j)\big((1-\dt_{j0})v_{j-i}(x)+(1-\dt_{i0})\hat{v}_{j-i+1}(x)\big)\dt'(x-y)
\nn\\
&\qquad\quad-(2j-1)\big((1-\dt_{j0})v_{j-i}'(x)+(1-\dt_{i0})\hat{v}_{j-i+1}'(x)\big)\dt(x-y).
\end{align}
\item[ii)] The second bracket: for $i,j\geq0$,
\begin{align}\label{}
&\{v_{-i}(x),v_{-j}(y)\}_2^{[0]} \nn\\
&\qquad=\sum_{r=-1}^{i-1}\Big(2(i+j-2r-1)v_{-r}(x)\,v_{-i-j+r+1}(x)\,\dt'(x-y)\nn\\
&\qquad\quad+(2j-2r-1)v_{-r}(x)\,v_{-i-j+r+1}'(x)\,\dt(x-y)\nn\\
&\qquad\quad+(2i-2r-1)v_{-r}'(x)\,v_{-i-j+r+1}(x)\,\dt(x-y)\Big) ,\\
&\{\hat{v}_{i}(x),\hat{v}_{j}(y)\}_2^{[0]}=-\sum_{r=0}^i\Big(
2(i+j-2r+1)\hat{v}_r(x)\,\hat{v}_{i+j-r+1}(x)\,\dt'(x-y) \nn\\
&\qquad\quad+ (2j-2r+1)\hat{v}_r(x)\,\hat{v}_{i+j-r+1}'(x)\,\dt(x-y)\nn\\
&\qquad\quad+(2i-2r+1)\hat{v}_r'(x)\,\hat{v}_{i+j-r+1}(x)\,\dt(x-y)\Big),\\
&\{v_{-i}(x),\hat{v}_{j}(y)\}_2^{[0]} \nn\\
&\qquad=\sum_{r=\max\{-1,i-j-1\}}^{i-1}\Big(
2(i-j)v_{-r}(x)\,\hat{v}_{-i+j+r+1}(x)\,\dt'(x-y)\nn\\
&\qquad\quad+(2r-2j+1)v_{-r}(x)\,\hat{v}_{-i+j+r+1}'(x)\,\dt(x-y)\nn\\
&\qquad\quad+(2r-2i+1)v_{-r}'(x)\,\hat{v}_{-i+j+r+1}(x)\,\dt(x-y)\Big).
\end{align}
\end{itemize}

\section{Concluding remarks}
Based on the Lax pair representation \eqref{PPht}, \eqref{PPhth} of
the two-component BKP hierarchy, we obtain a bihamiltonian structure
of this hierarchy. Our method in the construction of the Poisson
brackets is to employ the standard $R$-matrix formalism, which is
analogous to that for the two-dimensional Toda
hierarchy~\cite{Carlet}. In comparison with the two-dimensional Toda
hierarchy, we expect that there would be an infinite-dimensional
Frobenius manifold underlying the two-component BKP hierarchy.

As shown in \cite{LWZ}, the two-component BKP hierarchy
\eqref{PPht}, \eqref{PPhth} is reduced to the Drinfeld-Sokolov
hierarchy of type $(D_n^{(1)},c_0)$ under the constraint
$P^{2n-2}=\hat{P}^2$. Whether such a constraint induces a reduction
of the bihamiltonian structure is unclear yet. We hope that
considering this example would help to understand the relations
between Frobenius manifolds of infinite and finite dimensions.

\vskip 2ex

\noindent{\bf Acknowledgments.} The authors thank Si-Qi Liu and
Youjin Zhang for their advice, and thank Yang Shi for her comment.
They are also grateful to the referee for the helpful suggestions.
This work is partially supported by the National Basic Research
Program of China (973 Program) No.2007CB814800 and the NSFC
No.10801084.

\end{document}